# THE DIFFUSE LIGHT OF THE UNIVERSE
## On the microwave background before and after its discovery: open questions

Jean-Marc Bonnet-Bidaud[1]

**Abstract**

In 1965, the discovery of a new type of uniform radiation, located between radiowaves and infrared light, was accidental. Known today as Cosmic Microwave background (CMB), this diffuse radiation is commonly interpreted as a fossil light released in an early hot and dense universe and constitutes today the main 'pilar' of the big bang cosmology. Considerable efforts have been devoted to derive fundamental cosmological parameters from the characteristics of this radiation that led to a surprising universe that is shaped by at least three major unknown components: inflation, dark matter and dark energy. This is an important weakness of the present consensus cosmological model that justifies raising several questions on the CMB interpretation. Can we consider its cosmological nature as undisputable? Do other possible interpretations exist in the context of other cosmological theories or simply as a result of other physical mechanisms that could account for it? In an effort to questioning the validity of scientific hypotheses and the under-determination of theories compared to observations, we examine here the difficulties that still exist on the interpretation of this diffuse radiation and explore other proposed tracks to explain its origin. We discuss previous historical concepts of diffuse radiation before and after the CMB discovery and underline the limit of our present understanding.

[1] Astrophysics Department, French Alternative Energies and Atomic Energy Commission (CEA), 91191 Gif-sur-Yvette, France, E-mail : bonnetbidaud@cea.fr

# 1. Introduction

Modern cosmology, with objectives to give a scientific description of the universe, is a relatively young discipline. It is largely based on a series of accidental discoveries. In the beginning, there was of course an unexpected observation in 1929, by the American astronomer Edwin Hubble, which led to think that galaxies, islands of billions of stars, are all moving away from each other with speeds increasing with distance [1]. This apparent cosmic "expansion " ended the Albert Einstein dream of a universe in perfect equilibrium that was designed by him through a static universe model in 1917 [2]. It gave the impression instead that, in the past, the universe could have been hotter and denser. The hypothesis of this initial warm phase is the basis of all current universe models called "big bang" – in reference to a hypothetical brutal onset of the expansion – with the first version established in 1922 by the Russian physicist Alexander Friedman [3]. The cosmic expansion models have long had the failure to predict a younger universe than Earth, and therefore had only a limited echo until 1965.

At that time, an almost accidental observation put them back on the front of the stage. The discovery by two American engineers of a cosmic radiation, unnoticed until then in between radio waves and infrared radiation, brutally gave a new colour to the big bang hypothesis [4]. Indeed, this diffuse radiation that filled the whole sky can be interpreted naturally as remains of the hot phase. It corresponds to the first light produced in the hot universe, thereafter cooled by the dilution effect of the expansion. This first light, called also "fossil radiation " or "cosmic diffuse background (CMB)" has become today the Holy Grail of modern cosmology. Increasingly precise maps of the distribution of this primordial light were established and numerous publications deduce from the characteristics of this radiation only, all the cosmological parameters that govern the evolution of the universe. These results, which are often given with surprisingly reduced margins of uncertainty, even gave birth to the concept of " precision cosmology [5]".

The diffuse radiation is more than dominantly accepted as the direct evidence of the hot big bang. However strictly speaking, it was never part, of the big bang predictions, contrary to what is often claimed – and its accidental discovery is enough to prove this. So, is it really inseparable from the big bang? Can the same radiation be interpreted differently by other mechanisms and cosmological theories?

We examine here the strong and weak points of the present interpretation of the CMB. We will in particular examine the term often used of "cosmic radiation" which should be understood here as "related to the origin of the universe". Initially, the diffuse radiation was only a detection of a new radiation. It is only though its interpretation that we can later decide of its "cosmological" nature. This first step of the analysis is often quite widely eluded in current cosmological discussions, just as it was at the time of its discovery. To better understand the present status of this very specific observation, we must return to the time and circumstances of its discovery.

## 2. The discovery of the microwave diffuse radiation

The development of modern cosmology owes much to the Second World War and radar technique that was developed during the conflict. It is no coincidence that the majority of great post-war cosmologists, including English and Americans, all worked on this detection technique, using for the first time the high frequency radio waves. Observed from the ground, these waves were also the first new "window" in the observation of the universe, beyond visible light. It is with a radiometer derived from the radar technique, that in 1964, two scientists, Arno Penzias and Robert Wilson, from the Bell Telephone Laboratory (New Jersey, USA) were able to undertake this new kind of observations. With their radiometer, cooled with liquid helium and placed behind a large 6-m wide horn-shape metallic antenna, they were able to channel efficiently these radio waves. Their goal was to observe the sky in the hope of detecting very low radio emission that may come from the halo of our galaxy.

After numerous checks, they are faced with an embarrassing result. They detect very weak radio radiation but it covers the whole sky evenly. After rejecting several hypotheses, including that of an atmospheric disturbance related to the numerous contemporary tests of atomic weapons, they decide to publish their results in 1965 with a very cryptic and cautious title "*A measurement of excess antenna temperature at 4080 megacycles/s*" [4]. In particular, they determine that the radio radiation they detect at 7.3 cm wavelength corresponds to a very low temperature of only 3.5±1K (K for the Kelvin unit, or 3.5 degrees above absolute zero), without providing explanation on its actual origin. Yet, it is this very technical article, without any justification on the nature of the phenomenon that is now cited as one of the greatest proofs of the big bang model validity and incidentally allows its authors to receive the Nobel Prize in Physics in 1978.

The link with the Big Bang cosmological model was far from clear at the time. Shortly before publication, Penzias and Wilson had been warned that the team of Robert Dicke of Princeton University (USA) were developing the same type of instrument to determine the average temperature of the universe. At that time, Dicke sought by this measurement to check his own cosmological model, a cyclical universe, made of an infinite succession of expansion and contraction phases. Dicke, recognizing that he had been overtaken in this search, hastened to publish an article associated with that of Penzias-Wilson, in the idea of consolidating its own model [6].

The real relation of the diffuse radio radiation with the big bang was subsequently often associated with the name of George Gamow, but quite wrongly. The Russian physicist, a student of Friedman and later emigrated to the US, has been a staunch promoter of the idea of the Big Bang, including attempting to explain the formation of all chemical elements during the warm phase, but he was not the first to mention a residual big bang radiation. These are two of his collaborators, Ralph Alpher and Robert Hermann who first derive in 1948 that the big bang warm phase, after the expansion effect, should now have a residual temperature of 5K – value that they altered later to 28K [7,8]. Their publication in the Nature journal of November 13, 1948 was a very discreet and short note of only 50 lines which ended almost anecdotally by "*the gas temperature at the time of condensation (of the galaxies) was 600K and the temperature in the universe at the present time is about 5K*". It is this brief mention that is considered as "THE" big bang prediction. In reality, the assessment went completely unnoticed at that time in 1948, when the big bang model was still predicting an incoherent age of the universe. And in 1965, nobody remembered that the big bang could produce such radiation.[2]

## 3. The 3K radiation as a fossil radiation

Since its discovery, the diffuse microwave radiation, also often called 3K radiation, was very generally received as a "proof" for the Big bang, even though its temperature had not been actually predicted – various values ranging from 2 to 50 K had been put forward – and even if, as we shall see below, such a temperature can also be interpreted more generally as a temperature of the interstellar medium.

We place ourselves for the moment in the frame of the big bang where the origin of the radiation find however a very explicit interpretation. It results from a specific time of expansion where the density and temperature of the universe diminish enough so that light can escape from matter. At that time, generally evaluated between 300 000 to 400 000 years after the beginning of the expansion, the atom nuclei already formed (mainly hydrogen and helium) are surrounded by clouds of electrons which trap the light. When the temperature falls below a certain threshold, of about 3 000 K, electrons are captured by the nuclei leaving the way open to light. It is then considered that the particles of light, the photons, are definitely released and can cross the universe without experiencing any further significant alterations. In this sense, therefore, 3K radiation that reaches us today would be the first light produced in the universe and could be considered as "fossil" since it originates back to the early beginning of the expansion.

These circumstances imply very specific characteristics for the 3K radiation. Indeed, at the time when electrons are captured – a process also called recombination – light and matter are in equilibrium, with the light being equally likely to be emitted or absorbed. In this case, the light energy is spread according to a particular distribution called "blackbody" with reference to the theoretical arrangement of a black cavity in which light is perfectly absorbed and re-emitted. One of true big bang model predictions is therefore that the diffuse radiation must have a "blackbody" distribution. In 1965, Penzias and Wilson had obtained a measure only at one energy and thus could not confirm this prediction. This did not prevent the community of astrophysicists to immediately see a confirmation of the big bang and, at the same time, a rejection of other competing cosmological models. At that time, models such as the stationary universe of Fred Hoyle, Hermann Bondi and Thomas Gold, could not explicitly justify the existence of this radiation. Only a few years later, thanks to balloon flights in the upper atmosphere, it was possible to give more weight to the Big Bang hypothesis by capturing the background radiation at different energies. Then in 1990, the COBE satellite measured the overall radiation between the wavelengths of 1 to 100 cm and provides complete distribution of the light energy, in perfect agreement with a "blackbody" at a temperature of exactly $T = 2.725 \pm 0.002K$ [9]. This distribution is certainly today the strongest argument in favour of a warm phase as put forward in the big bang. This undeniable success earned George Smoot and John Mather, the two leaders of the COBE experiment, the Nobel Prize for Physics in 2006. Yet, it hides a major challenge for the big bang model and triggered a cascade of fundamental changes in the description of the expansion.

---

[2] Note that the discovery of the 3K radiation could have been attributed to a French team. Jean-François Denisse, James Lequeux and Emile Roux, three physicists at the Ecole Normale of Paris, observing with a German Würzburg radar antenna at the wavelength of 33cm, reported in a communication to the Academy of sciences on 17 June 1957, 7 years before Penzias and Wilson: *"Our measures have however shown that the sky brightness temperature is less than 3K and its variations from one point to another is less than 0,5 degree.* This is the first known measure of this famous diffuse radiation, though the authors did not realize the scope of their observation. See Denisse J.F., Lequeux J., Le Roux E. (1957) "Nouvelles observations du rayonnement du ciel sur la longueur d'onde 33 cm", CR Acad Sci 244, 3030.

## 4. The texture of the diffuse radiation

The COBE satellite in 1992 provided the first complete map of the diffuse radiation with sufficient resolution to discern details, including small temperature variations. Under the big bang hypothesis, the warm phase cannot be completely homogeneous. In order to subsequently form galaxies by the action of gravity, it is indeed necessary that there exist "lumps" in the primordial soup, relatively dense regions around which the matter will later aggregate. Since these inhomogeneities are kept in memory, imprinted in the primordial light, they should be found back in the form of temperature differences in the sky. Such temperature differences had been predicted in 1970, independently by the Russian astrophysicist Yakov Zeldovich and American Jim Peebles [10,11] without anyone being able to discover them. The COBE satellite was designed to catch these variations but it brought a surprise. Inhomogeneities do exist, but they were found 10-100 times lower than predicted. From one sky region to another, it was observed only infinitesimal variations of 0.001%, equivalent to temperature variations of only 30 millionths of degrees [12]! By analogy with the irregularities at the surface of the Earth, this represents the height of a molehill compared to the dimension of Mount Everest! As a result, corresponding initial lumps are too tenuous: with such uniformity, it is not possible to form galaxies in the limited time that has elapsed since the big bang. This serious anomaly was the source of a profound revision of the standard model of the Big Bang. To accelerate the condensation of matter, a huge amount of additional material should be added, different from the visible matter. This material, called "dark matter" had to be introduced in an amount ten times greater than the observed matter to provide the additional gravity needed to allow matter condensation. Yet, even if the indirect evidence of dark matter has been suggested from the movement of stars in galaxies and galaxies within clusters, its real existence has not been demonstrated up to date, despite over thirty years of tracking in laboratories.

The price to pay for this too perfect uniformity of the background radiation is therefore the introduction of an unknown that quickly leads to another one. Indeed, this major "dark matter" adjustment leads to an embarrassing effect on the age of the universe. An additional large amount of matter, by its higher gravity, significantly slows down the expansion, decreasing at the same time the age of the universe. The universe, if made of all the dark and visible matter, should then have no more than 9 billion years of existence. It would therefore be younger than some of its oldest stars! To balance this effect and allow a longer age of the universe, it is then essential to add a repellent action that could oppose the increase effect of gravity. This analysis led cosmologists to introduce the concept of "dark" energy, representing this time by itself, about two times more energy than all the matter in the universe. So, it is no coincidence that these two " joker" notions of dark matter and dark energy appeared in the 1990s. They have become essential to bypass the anomaly brought by an excessive smooth texture of the background radiation, if this radiation was produced in the big bang. Later, in 1998, indication of an accelerated expansion was suggested from supernovae at great distances, and provides a rationale to justify the hypothesis of dark energy.

It is on this new basis, with these two additional unknown components of dark matter and dark energy, that the diffuse radiation is interpreted today. It should be noted that, in the big bang, the granularity of the first light is not actually predicted by the cosmological model since its initial conditions are not known. To get the currently observed irregularity, it is necessary to introduce arbitrarily, "by hand", some selected initial density fluctuations, very early in the expansion. These will then develop in matter until recombination time when they remain frozen in the first light. Their apparent size observed today on the sky depends therefore not only on these arbitrary initial parameters but also, by a projection effect due to the subsequent expansion, on the set of parameters of the universe - density, proportion of dark matter, dark energy, etc ... No fewer than eleven independent parameters are necessary to describe them today.

The WMAP satellite, into orbit from 2001 to 2008, examined with a much more powerful "magnifying glass" than COBE the dimensions of 3K irregularities, down for the first time to sizes less than the degree. The main result was the discovery that the distribution of inhomogeneities by size is not uniform. Some sizes appear much more often than others and, if we establish a count of these inhomogeneities according to their size, the curve shows several peaks. The majority of the fluctuations have sizes of about 1° on the sky, defining the position of a first peak. A second group of fluctuations have sizes around 0.3° and optionally a third group still poorly determined around 0.2 ° [13]. These different peaks of fluctuations each provide specific information. The position of the first peak mainly depends on the total density of the universe and fixes its geometry, as substantially flat. The peak amplitudes also constrain the amount of dark and visible matter in the universe.

By a trial and error method, it is then possible to determine a series of cosmological parameters that reproduce accurately the peak fluctuations. The level of detail in which was conducted this analysis has now resulted in an impressive list of 30 parameters of the universe whose values can be determined through these adjustments with very low uncertainties. For example, the age of the universe was calculated to be 13.74±0.11 billion years, with uncertainty barely over 100 million years, and the proportion of dark matter was estimated to be 23±2% of the universe critical density [14]. Note that these uncertainties are formal, reflecting only statistical effects. They merely reflect the possible differences in the context of very specific assumptions and ignore multiple possible parasitic effects that may occur in the reduction process, such as calibration errors or bad subtraction of local inhomogeneities produced in the Galaxy.

The European Planck satellite, collecting data from August 2009 to October 2013, recently obtained data of higher quality with better spatial resolution and further refined these results. In the first release in 2013, the age of the universe was originally set to 13.82±0.06 billion years and the proportion of dark matter 26.3±0.7%, later revised in the full final release of February 2015 to 13.813±0.038 billion years and 26.4±0.5% [15,16]. Apart from what may appears as a pure numerology play, the main Planck contribution was to provide information on the polarized component of the 3K radiation, that is properties of light in different planes. In principle, this enables to search for promising tenuous imprints of what may have been the conditions prior to recombination that produces the 3K. For instance, this gives potentially access to the hypothetical gravity waves that might have accompanied the very brutal and dizzying accelerated expansion of the early universe called "inflation". The discovery of these gravity waves from polarized light was claimed by a ground-based experiment Bicep-2 but finally denied using the more complete information collected by Planck [17,18].

Behind the apparent convergence of some basic cosmological numbers and impressive amount of information derived from the 3K, a few major discrepancies remain however hidden that the CMB community has chosen to label with the mild term of "tension". The most important "tension" concerns the major cosmic parameter, the Hubble constant, Ho, which quantifies the universe rate of expansion. The Planck value, if fitted to describe the observed 3K fluctuations, is estimated at Ho=67.31±0.96 km/s/Mpc [16] while the value carefully measured from the motion of local galaxies by the Hubble telescope Key project yields Ho=73.8±2.4 km/s/Mpc [19]. Taken at face value, the expansion rate that allows to produce the 3K fluctuations at the right amplitudes do not fully agree with the astronomical measurements. Other concerns are related to the amount of dark matter and its effect on the abundance of large galaxy clusters. Given the large amount of suspected dark matter (26%), about twice as much galaxy clusters are expected to form compared to what is actually observed. This "tension" could only be released if part of the dark matter is very specific, for instance in form of a "light" neutrino [20].

These discrepancies are so far considered as minor and, if you believe some experts like Wayne Hu of the University of Chicago, the "cosmological" interpretation of the microwave background should soon lead us to a situation where " *we will then know as much, or even more, about the early Universe and its contents as we do about the fundamental constituents of matter* [21]"!

.

A majority of astrophysicists probably shares this beautiful enthusiasm and seems to consider as definitely established these results derived from a unique radiation and many assumptions. Yet the history of science teaches us that what seems demonstrated one day may be challenged the next day. As such, the unfortunate story of physicist William Thompson, inventor of the second law of thermodynamics and the concept of absolute zero, is a cruel illustration. In the late XIX$^e$ century, this prestigious scientist had determined the age of the sun as 100 million years on the basis of its supposed cooling time and had imprudently stated "*There is nothing new to be discovered in physics now, just to make measurements more accurate. The future of physics is now to determine the sixth decimal place* " [22].

He was just unable to imagine that the real source of nuclear energy from the Sun would be revealed just a few decades later, by the advent of relativity and quantum mechanics. Today, cosmology could have reached the same crossroads. Who would have bet 20 years ago, on a universe dominated by dark matter and dark energy? With such present unknowns, should we really believe that it is enough to simply adjust the last decimal places? There is obviously some success in the current description but is it the only possible conclusion? Is there still room for other explanations and interpretations of the diffuse radiation? Since the famous 1965 discovery, many scientists have tried other approaches, based on alternative assumptions partially or totally different from the big bang.

## 5. A universe without dark matter

An alternative, perhaps the more developed today, was born at the beginning to solve the riddle of unfound dark matter. It is based on the idea that the existence of dark matter might be just an illusion. In fact, dark matter intervenes everywhere where gravity must be high and where visible matter is limited. Two solutions are possible; either to increase the amount of matter or increase the gravity. It is this second path that was followed by astrophysicist Mordechai Milgrom in 1983 by simply supposing that the law of attraction between masses was slightly different than the one we tested in the laboratory and in the solar system [23].

Beyond a certain distance, the attraction between two masses would decrease more slowly than predicted by the law of universal gravitation. At great distances, gravitation would then be stronger than expected. This simple "cosmetic" change of gravitation, called MOND (for MOdified Newtonian Dynamics), is sufficient to explain the anomalies in the mass of galaxies. Despite its success, this assumption has long been neglected because it seemed difficult to reconcile with general relativity. But in 2004, the Israeli physicist Jacob Bekenstein managed to generalize MOND to make it compatible with the theory of Einstein at the cost of the introduction of three additional parameters, in an enlarged theory dubbed "MOND-TeVeS " [24].

This theory, which now appears as a complete theory of gravitation, can also account for the gravitational lenses, these small image distortions normally also attributed to dark matter. It is also able to predict the early evolution of the universe that gave rise to initial matter lumps and light fluctuations [25,26]. From the same initial conditions as the conventional big bang, the universe of MOND-TeVeS, with a different gravity and therefore without dark matter, generates a very similar diffuse background but with characteristics that are determined by very different parameters. Surprisingly for a theory that was not built for this purpose, it is able to reproduce faithfully enough the size distribution of temperature fluctuations, including the first and second peaks. Differences with the prediction of the classic big bang is now concentrated on the third peak of the fluctuations. To account for this, MOND-TeVeS may require the introduction of massive neutrinos [27,28]. These particles, long regarded as massless, are now strongly suspected of having a low but non-zero mass that could not yet be measured. MOND may therefore also rely on some unknown component to provide a CMB description as good as the big bang with its unknown dark matter and dark energy.[3]

## 6. The temperature of space

Other explanations of the 3 K radiation arise from a more fundamental question about the exact distance where this cosmic light is formed. For estimating a distance, astronomers assume that the expansion is uniform and make use of the redshift of light, generally by measuring wavelength markers corresponding to particular chemical elements. But nothing of the sort is visible in the background radiation as its blackbody distribution is particularly smooth. No direct distance measurement can therefore be performed. At a consequence, this radiation could actually be produced - for all or part - in any region of the universe. According to its distance, the deduced actual temperature varies because there is a simple relationship between temperature and the redshift, if it is attributed to an expansion. The farther is the radiation, the more the temperature will seem weakened. The 3K diffuse radiation that reaches us today may therefore correspond as well to radiation at 3000K issued at a redshift of 1000 (this is the assumption for the fossil CMB) or radiation emitted at 300K at a redshift of 100 or obviously also radiation emitted at 3K at a redshift 0 (ie locally), with of course all possible intermediate situations.

---

[3] Concerning MOND, see also the critics based on the observations of colliding clusters (Clowe et al. (2006) « *A Direct Empirical Proof of the Existence of Dark Matter* »,ApJ. 648, L109) and the rebuttal by Mc Gaugh (http://astroweb.case.edu/ssm/mond/bullet_comments.html)

This aspect greatly complicates the interpretation of the phenomenon. Indeed, there is presently only one indirect indication of the possible large distance of the background radiation. It is a subtle effect of 3K temperature distortions due to hot gas inside galaxy clusters. This effect, called Sunyaev-Zel'dovich (or SZ, named after two Soviet astrophysicists who had predicted its existence in 1969 [29]), is actively sought and studied. First extensive catalogs of SZ clusters were published in 2015, following Planck observations. The SZ effect indicates that at least part of 3K comes from beyond some galaxy clusters. but its exact interpretation is very delicate since it depends on the characteristics of each cluster and it mixes with other temperature distorsions.

The nature of diffuse radiation is therefore a much more open question than it seems a priori. Is it really only a very distant fossil background radiation or simply a more universal radiation, filling the entire space and produced both locally and at long distances?

This problem has been addressed long before the 3K discovery, almost at the beginnings of modern astrophysics when the question arose of the temperature of space. Indeed, we know that the universe is filled with radiation coming from all cosmic objects. So what would be the measured temperature if we put a thermometer in space? In other words, what will be the temperature of all the radiations of the universe if they are transformed into heat? This is to the Swiss Charles Guillaume that we owe the first estimate of this temperature in 1896 [30]. This physicist, who headed the International Bureau of Weights and Measures in Paris from 1915 to 1936, was awarded the Nobel in 1920 for his discovery of low expansion alloys. He considered the effect of the radiation of all the stars on a given point in space. His estimate was necessarily an approximation of which he was very aware at that time, writing *"when trying to calculate the temperature increase produced in a point in space by the radiation of the stars, a great difficulty is encountered in the evaluation of the energy they radiate."*[30]

Nevertheless the calculation, even approximate, gives a surprising result: the inferred temperature is 5.6K. English astrophysicist Arthur Eddington improved the same calculation later in 1926 and found a fairly close result of 3.2 K [31]. Thus, locally, the temperature of space under the action of the radiation of the stars is approximately the same as the temperature of the microwave background radiation.

The same conclusion was drawn later after the discovery of cosmic rays, this high speed flow of particles that cross our galaxy. German physicist Erich Regener calculated in the same way in 1933 the temperature rise due to cosmic rays and deduced an equilibrium temperature of 2.8 K [32]. The similarity of all these temperatures with each others and with that of the microwave background radiation is perhaps no coincidence. It may reflect the fact that the universe is in balance, each of these components having had time to equilibrate. The light radiation of the stars is of course a local characteristic linked to galaxies but the cosmic rays radiation is a more universal component since these high-speed particles are also present in intergalactic space. In other words, it seems rather easy to produce everywhere in the universe a "universal" temperature of a few degrees above absolute zero. It is this reasoning that guided several scientists to propose an alternative explanation for the 3K radiation.

## 7. The return of the stationary universe

The first to grasp the importance of the space temperature concept were supporters of the Steady-State Theory, the competing cosmological theory of the big bang, developed in 1948 by the trio of astrophysicists Hoyle, Bondi and Gold [33]. In this theory, the universe was not created, it is eternal, and in this sense it respects the "perfect cosmological principle": a universe identical at any point in any direction and in any moment. The expansion is produced by a very low and continuous creation of matter. This theory did not explicitly forecast a diffuse radiation at the time of the 3K discovery, a deficiency that makes it lose some of its credit. But its authors have not fallen pavilion. They soon argued that for the big bang as well, neither the temperature nor the intensity of radiation 3K were really predicted.

The microwave diffuse radiation is a radiation of very low intensity. Received on Earth, it represents only 3 microwatts per square meter, which is 300 million times lower than the energy received from the sun. This corresponds to a density of approximately 400 photons for each cubic centimetre of space. Given their wavelength, these photons participate also at least partially to the "snow" seen in our TV sets: they are part of this background noise that fills the screen when no strong signal is detected! In the universe, it's a very low radiation density that represents in energy only a tiny fraction, about 0.006% of the average energy density of the universe, compared to 4% for example for ordinary matter in the big bang. Moreover, in this cosmological interpretation, the intensity of the radiation is totally arbitrary. It is not calculated or calculable a priori from the earlier expansion characteristics. On the contrary, it is its measure that conversely provides an estimate of the number of photons in the universe, a data necessary to calculate the formation of light elements in the early universe. In the big bang, the background radiation could be just as well a thousand times weaker or stronger. It can only be constrained by further independent assumptions on the abundances of very few light elements. The coincidence of its intensity and its temperature with that of other existing types of radiation thus appears fully coincidental in the big bang context.

Invoking the concept of space temperature, Hoyle, Geoffrey Burbidge and Jayant Narlikar argued on the contrary that this coincidence could be explained more naturally [34]. It just requests to consider the background radiation as the consequence of another general process in the universe: the conversion of hydrogen into helium by thermonuclear reactions inside stars. Burbidge and Hoyle were well placed on this topic because they were the ones who provided in 1957 the first detailed calculations to estimate the production of elements in the universe [35] and, by 1964, Hoyle had precisely completed these calculations for helium [36]. The result is amazing: the amount of energy produced by the fusion of hydrogen to produce all the helium observed in the universe is exactly the one contained in the diffuse background. Calculating the temperature of space corresponding to this energy, gave them a typical temperature of 2,68K [34, 37]. The result appears so close to the 3K radiation that for Hoyle, it cannot be a simple coincidence. However, a problem remains to solve because the energy released by the fusion of hydrogen in helium is in the form of visible or ultraviolet radiations, not microwave radiation. Nevertheless, there are several processes that can degrade a radiation into another. The one considered by Hoyle is the diffusion of light through cosmic dust. The radiation is then degraded from visible to infrared, as it is the case through the atmosphere when the sun sets down. The innovation designed by Hoyle is to consider that a fraction of this dust is made of wiskers, fine carbon and iron needles [38]. This type of dust may be released in large quantities by the explosion of massive stars and some laboratory experiments have shown that the dust may then not condense into fine spherical grains but into needles of about 20 nanometers in diameter and 1 millimeter long. The peculiarity of these needles is to turn the light very efficiently into ... microwave radiation. Close to the stars, there exists therefore both a natural source of light, produced when the stars burn hydrogen into helium, and also a natural diffusing material, dust made of needles, which can produce microwave radiation with the intensity and temperature observed over the entire sky [37].

One of the main problems faced with such process is however the difficulty to produce a pure Planckian spectrum as observed. This requires a perfect thermal equilibrium and, till more precise information is collected on the dust characteristics and diffusion properties, no precise information can be derived. It also remains to solve the almost perfect uniformity of the radiation through space, with however tiny temperature fluctuations. Since its creation, the steady state theory has evolved to better account for the creation of matter needed to maintain the expansion. It now considers a "quasi-stationary universe": the universe would be in perpetual oscillation with a contraction-expansion cycle lasting about 40 billion years, superimposed on a general expansion lasting since at least 1000 billion years (hence the name of quasi-steady state theory) [34]. Rather than a continuous process difficult to explain, the matter is now created only in these phases of contraction, especially in high gravity regions, such as those close to massive black holes In addition to allowing natural diffusion by carbon and iron needles, these phases of contraction contribute to make the radiation uniform very efficiently. As for the small fluctuations, they simply correspond to the excess of diffusion generated by the concentration of matter in clusters and groups of galaxies. In 2007, by simply introducing these effects, Narlikar and Burbidge - the two co-authors of the quasi-stationary universe model with Hoyle now deceased - have largely been able to reproduce results from WMAP on the size of fluctuations [39]. Actually, in this exercise, the model of quasi-stationary universe enjoys great freedom because inhomogeneities of widely varying sizes may appear simply as the consequence of the large scale distribution of dust. In the interpretation of Hoyle and his collaborators, the background radiation could be then one of the best ways to locate the dust in the universe. They also argued that the existence of this dust could significantly alter the measurements of star explosions, the supernovae, which are used today to measure the rate of expansion. Some of these explosions appear fainter than expected and this led to the idea of an accelerated expansion of the universe, associated today with dark energy. But much of this effect could only be due to simple absorption by dust. To give more strength to the Hoyle interpretation, this specific type of dust should be detected directly. This has not yet been achieved so far.

The quasi-steady state cosmological model offers an example of a diffuse microwave radiation, not primordial as in the case of the big bang, but rather universal because created throughout space by known general physical processes. It has also the merit to predict more naturally the intensity and temperature of the radiation. Contrary to early conclusions drawn in 1965, such a cosmology can therefore explain the background radiation, although it was not explicitly predicted initially. The background radiation is therefore not really a verdict between two rival theories. Yet, the stationary universe cosmology is no longer supported by most astrophysicists because of uncertainty about the actual process of matter creation on which it relies. It is certainly less developed than the big bang but the arguments are more balanced than it seems. And it was easy for Fred Hoyle to object with his rebellious eloquence *"the 3K prediction by the big bang is a scam. The big bang never predicted the temperature of the microwave radiation supposedly cosmological. It depends entirely on initial conditions"* adding that for cosmic matter "*matter creation has never been a problem in our theory because it occurs naturally as the creation of matter-antimatter pairs, unlike the conventional big bang where the 'initial creation' violates all laws of invariance*"[40].

This longstanding debate between Hoyle "stationary" universe and big bang "creationist" universes was therefore not fully resolved by the discovery of background radiation. It will eventually find its epilogue when clear indications will be obtained on undisputable changes in the universe with redshift. The big bang indeed predicts that, at high redshifts, the universe must be totally different, with stars and galaxies in the early stages of their evolution. For now, our observations at very large distances are still too limited to provide truly compelling evidence of these evolutionary effects.

## 8. The plasma universe

Production of microwave radiation by the diffusion of a local radiation was also invoked in other cosmological models such as the plasma universe, originally developed in 1965 by the Swedish Hannes Alfven, Nobel Prize in Physics in 1970 for his work on plasmas [41,42]. This cosmology questions another fundamental assumption of the big bang: is it legitimate to consider that gravity is the only force shaping the evolution of the universe? To this question, Alfven answered negatively. For him, in fact, matter in the universe is composed almost exclusively by electrically charged particles and it is therefore logical to imagine that the electromagnetic force, the only other force that shares with gravitation an infinite range, can play a decisive role. For Alfven, the observed structure of the universe into vast filaments of matter was an evidence of this, because the action of the magnetic field produced naturally jets and linear structures in filaments. He confided, *" I have never thought that you could obtain the extremely clumpy, heterogeneous universe we have today, strongly affected by plasma processes, from the smooth, homogeneous one of the Big Bang, dominated by gravitation"* [43].

His cosmology imagines a hierarchical universe in which all galaxies form from a "Metagalaxy". From great distance, this metagalaxy is seen as a uniform background of galaxies and the light radiation produced by the stars, is then scattered by the many existing plasma filaments, so that it appears in the form of diffuse microwave background. Subsequent work by Eric Lerner showed in 1995 that these plasma filaments could produce inhomogeneities as those initially observed by the COBE satellite [44]. But these studies have so far not been applied to the latest and more detailed results from WMAP and Planck. Maybe because the plasma cosmology, which postulates an eternal universe, is currently being studied only by a very small group of researchers. Yet, it meets one of the enigmas posed by the big bang, the fate of antimatter. This cosmology involves indeed a universe where matter and antimatter are present in equal quantity and grouped into two separate plasma bubbles, forming an "ambiplasma". The contact of these two regions is expected to produce a strong high-energy radiation that has never been observed, but the contact region might well be outside our "horizon" of observation and therefore remains inaccessible.

### 9. Alternative cosmologies

To produce the microwave background radiation, others various proposals were also discussed. In an article posted on April 2010, two Spanish astrophysicists Antonio Alfonso-Faus and Marius Fullana, considered that this radiation could be produced by the evaporation of primordial black mini-holes [45]. These microscopic black holes are believed to form in large numbers in the early universe and could remain invisible. They may theoretically constitute part or all of the dark matter. By slow evaporation, introduced by the English astrophysicist Hawking, they can however emit a weak black body radiation at a wavelength (and temperature) fixed only by their mass. Alfonso and Fullana computed that to produce a microwave radiation with the observed temperature of 2,72K, such mini black holes should have a mass equal to 0.7% of the Earth mass, that is the mass of a planet intermediary between Pluto and Mercury, but confined within a radius of only 0.06 millimetres. If they really exist with this specific mass, then the microwave diffuse radiation may be produced not by light released by atoms but by the effect of evaporation of this multitude of mini-black holes distributed uniformly. However, significant doubt remains about the exact intensity of the radiation emitted by evaporation. Current calculations show that this seems inadequate to explain the intensity of the observed microwave radiation.

Other researchers, such as astrophysicist André Assis, proposed that the background radiation may simply correspond to emission by matter at an equilibrium temperature of 2,7K, in the frame of a cosmology without expansion, where the redshift is interpreted as a "tired light" effect, that is to say a light loss of energy on its path [46]. In the same line, in the context of a quasi-static universe, Fahr and Zoennchen also proposed that starlight photons could become thermalized into CMB-photons through a photon-photon scattering mechanism [47]. Finally, to give a last example, in the "Curvature cosmology" proposed in 2006 by David Crawford, where the universe is dominated by a static hot plasma at high temperature, 3 K radiation is produced by diffusion on matter of high energy electrons into the plasma [48]. Note however that most of these works are currently too underdeveloped, particularly with regard to the explanation of the details of small inhomogeneities of the background radiation, but they illustrate the fact that there are multiple paths that can be followed to interpret this mysterious 3K radiation.

## 10. Strengths and weaknesses of diffuse radiation

Considering these different works, can we really consider the diffuse microwave radiation as an "indisputable" proof of the big bang? At minimum, it can be said that the actual situation seems much more mitigated. The alternative proposals certainly invoke specific hypotheses such as modified gravity, iron and carbon needles, plasma filaments or primordial black holes, as many components whose existence has not yet been demonstrated. But the big bang, in its current version also requires completely hypothetical processes. Admittedly, the process behind the formation of 3 K radiation, the light-matter interaction, is well known, but the detailed features of this radiation requires a succession of unknown mechanisms: the arbitrary initial conditions can only be justified by a hypothetical early inflation, this is followed by the need of an unknown dark matter to amplify the inhomogeneities and finally a still unexplained dark energy have to exist to accelerate the expansion. None of these components has yet received confirmation. So it could be said that the current interpretation of the big bang introduces more problems than it solves.

The interpretation of the background radiation as "cosmological" on the other hand has already its own contradictions, as revealed by the detailed analysis of WMAP and Planck data. First, there is the existence of hot and cold opposite points on the sky, defining a preferred axis, called "Axis of Devil" by astrophysicist Joao Magueijo [49]. Hopes were that increased resolution provided by the Planck satellite will cancel this defect but it was not so. The most detailed CMB maps obtained so far confirm the anomaly and point also toward a asymmetry between the North and South hemisphere [50]. This could be due to an incomplete removal of dust in the Solar system but it will then cast doubt on the overall validity of the CMB results. Until now, nobody has been able to explain this alignment except to imagine that the universe is not symmetrical, as seen from our point of view, which would undermine the cosmological principle on which is based the standard model of the Big Bang

In the diffuse background was also discovered and confirmed a very large area of about 10° on the sky, called "cold spot", where the temperature is abnormally low [50]. If this region is located at the assumed distance of the cosmic microwave background, it has gigantic dimensions and implies a inexplicable huge hole in the background of the universe. Statistically, it cannot be produced by the standard big bang with the supposed initial conditions of inflation and there is no known cosmic structure that can explain it. Some astrophysicists are now led to propose that it could result from "cosmic defects" predicted by string theory, or even from a parallel universe.

Additionally, a complete thermodynamic equilibrium between particles and photons at the time of recombination can be questioned and it was pointed out that, after the recombination phase, any small deviations from a pure unperturbed homogeneous expansion would unavoidably lead to deviations from the perfect Planckian blackbody spectrum [51]. Moreover, in the cosmological CMB interpretation, the temperature is expected to increase back in time and the hot radiation may efficiently prevent the collapse of cold molecular clouds into stars as early as for a redshift of 2 [51].

Overall, the interpretation of the microwave radiation is much more delicate and complex than it seems. If it is of cosmological nature and thus produced at very large distance, it is affected both by all the perturbations that light can undergo on its path but also by all different emissions that can be superimposed. In the range of microwaves, it already exists in our galaxy at least three sources of contaminating light: the local dust, the "synchrotron" emission produced when electrons circle along magnetic field lines and the "bremsstrahlung" emission when these electrons interact with nuclei of atoms. These contribution varies greatly with wavelengths: it is only 10% of the 3K background radiation at low wavelengths, but can reach two to three times the 3K signal intensity at higher wavelengths. The removal of this contaminating light is therefore a real headache and the 3K signal is indeed like a needle in a haystack. Doubts still persist on the overall validity of the operation, especially as it may also be plagued by errors in the reconstruction of satellite pointings and the time of arrival data such as reported for WMAP [52,53,54].

The current interpretation of 3K remains very complex and systematic uncertainties cannot be excluded, making probably still largely exaggerated the claim of "precision cosmology" that is presently drawn from it.

**11. The diagnostics of the future**

Even after the recent results of the Planck satellite, there is still a pending fundamental question about the 3K radiation: is it, yes or no, a distant background radiation, coming from a very early period in time?

To answer this question, few methods exist. We should be able, for example, to measure the temperature of the microwave radiation at different redshifts that is in regions more and more distant in the universe. We would then be able to check if, as predicted by the big bang, the further back we are in the past, the higher is the temperature. Apart by the microwave radiation, the only indirect measurement of 3K temperature used hitherto was to observe the cyanide molecule (CN). One of its transitions corresponds to energy very close to that of background radiation so that the molecule will be excited if bathed into such low temperature radiation. And, in retrospect, one could say that the background radiation had been already discovered in ...1941. At that date, the British Mc Kellar had already noted that, to explain the transitions observed from the CN molecule, it has to be immersed in a uniform 2.3 K radiation [55]! But currently, recent measurements show that the CN excitation temperature is in reality higher, by about 0,58K than the 3K background radiation [56]. This implies that most probably other more complex interactions are involved besides the 3K.

Other attempts to measure the temperature of 3 K radiation at different distances have been made using the transitions of the carbon atoms present on the line of sight in the direction of bright distant radio sources, the quasars. In 2000, with the largest VLT telescopes in Chile, a small temperature variation with distance has been measured, showing that the diffuse microwave temperature could be 2 to 5 times higher in the direction and close to a quasar with an estimated distance of approximately eleven billion light-years (corresponding to the light travel time estimated from Planck cosmological parameters) [57]. Similar studies were also produced using the CO molecules [58]. These results are so far the only indications of one of the most important predictions of the big bang: the increase in background radiation temperature with distance. But, as for the CN molecule, excitation of the carbon atoms or CO molecules may equally be produced, wholly or partly, by other processes than the microwave radiation, in particular by collisions of atoms or by stellar ultraviolet radiation. So these single results are still far from delivering the expected definitive proof. There is therefore no fully reliable thermometer of cosmic microwave radiation today.

A much more ambitious test would be to measure the redshift at the exact time when the 3K was produced that is to say at a redshift of approximately 1000. It seems that in the background radiation spectrum, small distortions may exist in the so-called Lyman-$\alpha$ hydrogen line, whose wavelength, and thus the redshift could be measured [59]. But this Lyman-$\alpha$ emission is largely "diluted" by other cosmic contributions and it may even not be within reach of the future largest radio telescopes, such as LOFAR, which is a network of thousands of antennas currently nearing completion in Europe [60] or SKA (Square Kilometre Array), which is a future international radio antennas project that will cover one million square meters [61].

Other hopes to prove the primordial warm phase imagined by the big bang can come from the study of neutrinos, these particles produced massively in nuclear reactions that interact very weakly with matter. In the Big Bang scenario, during the phase that immediately preceded the recombination producing the 3K, a wave of neutrinos was released after the formation of the first particles. As for the light, this should also now appear as a general diffuse background [62]. The temperature of this neutrino background is predicted to be about 2K. But, because of the difficulty to capture neutrinos and especially the huge background of neutrinos also produced locally, including in our nuclear power plants, the hope of being able one day to measure this temperature is very tenuous.

## 12. Conclusion

Without these fundamental checks, the "cosmological" nature of the background 3K radiation cannot be considered as fully demonstrated today. The 3K does not represent a major component of the cosmos and, contrary to a current consensus, diagnoses derived from it are less decisive than it seems. Due to its very low energy, it can be produced by a wide variety of physical processes. Its overall homogeneity and low level fluctuations can also be explained by different scenarios. Besides, its interpretation in the frame of the big bang brings insurmountable problems, because its too high homogeneity requires the introduction of enigmatic additional components such as inflation, dark matter and dark energy. These unknown ingredients are among the weaknesses of this cosmological model, as long as no direct indication of their existence is found.

So, we can only wonder at the ease with which a majority of scientists continue to present the particular scenario of the big bang as practically definitively established and refuse to give any credit to the various efforts to propose alternative ideas. An obvious dose of humility should instead be required in front of the evident limitations of our current observations. If our exploration of the local universe undeniably improves, we still know almost nothing of the objects that populate the universe at distances more than ten billion light years and, apart in visible light, the images we have at some other energies are often inferior to images available to Galileo with his telescope. As an example, the largest gamma ray telescopes currently in orbit has such a blurred vision that a perfect point of light in space at these energies appears through them as a spot the size of the sun. In addition, we know very little of a large part of the very high energy radiations that are still totally out of reach of our instruments. As the Chinese saying goes, *" The frog at the bottom of a well measures the extension of the sky with respect to the border of the well."* Our exploration of the universe is still very limited today and it remains possible that, due to unforeseen future discoveries, we will be led to consider in a few decades that our current efforts to decode the background radiation were as vain as to read coffee grounds.